# An Efficient Data Structure for Fast Mining High Utility Itemsets

Zhi-Hong Deng, Shulei Ma, He Liu

Key Laboratory of Machine Perception (Ministry of Education), School of Electronics Engineering and Computer Science, Peking University, Beijing 100871, China

**Abstract:** High utility itemset mining has emerged to be an important research issue in data mining since it has a wide range of real life applications. Although a number of algorithms have been proposed in recent years, there seems to be still a lack of efficient algorithms since these algorithms suffer from either the problem of low efficiency of calculating candidates' utilities or the problem of generating huge number of candidates. In this paper, we propose a novel data structure called PUN-list (PU-tree-Node list), which maintains both the utility information about an itemset and utility upper bound for facilitating the processing of mining high utility itemsets. Based on PUN-lists, we present a method, called MIP (**M**ining **h**igh ut**i**lity **I**temset using **P**UN-Lists), for fast mining high utility itemsets. The efficiency of MIP is achieved with three techniques. First, itemsets are represented by a highly condensed data structure, PUN-list, which avoids costly, repeatedly utility computation. Second, the utility of an itemset can be efficiently calculated by scanning the PUN-list of the itemset and the PUN-lists of long itemsets can be fast constructed by the PUN-lists of short itemsets. Third, by employing the utility upper bound lying in the PUN-lists as the pruning strategy, MIP directly discovers high utility itemsets from the search space, called set-enumeration tree, without generating numerous candidates. Extensive experiments on various synthetic and real datasets show that PUN-list is very effective since MIP is at least an order of magnitude faster than recently reported state-of-the-art algorithms on average.

## 1. INTRODUCTION

Frequent itemset mining, first proposed by [Agrawal et al. 1993], is an important topic in data mining. It plays an essential role in many important data mining tasks such as mining associations, correlations, causality, sequential itemsets, episodes, multi-dimensional itemsets, max-itemsets, partial periodicity, and emerging itemsets [Han et al. 2000]. In frequent itemset mining, an itemset is frequent if its occurrence frequency in a database is not less than a given threshold. Under this framework, the downward closure property [Agrawal et al. 1994] plays a fundamental role in various kinds of algorithms designed to discover frequent itemsets [Agrawal et al. 1994; Deng et al. 2010; Deng et al. 2012; Grahne et al. 2005; Han et al. 2000; J. Pei et al. 2001; Zaki et al. 2003].

However, frequent itemset mining does not take into account the unit profits and purchased quantities of items. Therefore, it cannot satisfy the requirement of the user who is interested in discovering the itemsets with high sales profits [Tseng et al. 2010]. This brings about a new interesting topic, which is called utility mining. In utility mining, each item has a utility value and can occur more than once in a transaction. Depending on the application, the utility of an item may be measured by price, profit, cost, etc. An itemset is called high utility itemset if its utility is not less than a given minimum utility threshold. Like frequent itemsets, high utility itemsets are considered valuable or interesting in many applications, such as business promotion and cross-marketing [Ahmed et al. 2009; Erwin et al. 2008; Tseng et al. 2010; Yen et al. 2007], website log and click stream analysis [Ahmed et al. 2010; H. F. Li et al. 2008; Shie et al. 2010; Yin et al. 2012], mobile commerce environment planning [Shie et al. 2011], pattern mining in bioinformatics [Chan et al. 2003], episode discovery in complex event sequences [Wu et al. 2013], and so on. The task of utility mining is to discover all high utility itemsets.

High utility itemset mining is much more challenging than frequent itemset mining because the downward closure property of frequent itemsets no longer holds for high utility itemsets. To address this mining task, lots of algorithms [Ahmed et al. 2009; Ahmed et al. 2011; Chan et al. 2003; Erwin et al. 2008; Lan et al. 2014; Y. C. Li et al. 2008; Y. Liu et al. 2005; M. Liu et al. 2012; J. Liu et al. 2012; Tseng et al. 2013; Tseng et al. 2010; Yao et al. 2004] have been proposed. Most of these algorithms work in the candidate-generation framework. The candidate-generation framework consists of two phases. In the first phase, candidate high utility itemsets, such as HTWUIs [Y. Liu et al. 2005], are found from a database. In the second phase, high utility itemsets are identified from the candidates by calculating the exact utility of each candidate. The main problem of the candidate-generation framework is that the number of candidates may be huge, which leads to bad scalability and low efficiency. To address this problem, a new framework, which discovers high utility itemset without generating candidates, is proposed in recent studies [M. Liu et al. 2012; J. Liu et al. 2012]. In the new framework, high utility itemsets are directly identified from a set-enumeration tree. Substantial experiments show that the new framework has better performance than the Two-Phase framework. However, the algorithms based on the new framework suffer from the problem of how to compute the utilities of itemsets efficiently.

In this paper, we address the above problem by proposing a novel data structure to facilitate the task of high utility itemset mining. The major contributions of this paper are as follows:

(1) A novel structure, called PUN-list, is proposed. The PUN-list of an itemset maintains both the utility information and the utility upper bound of the itemset, which acts as a pruning strategy to narrow the search space. In fact, PUN-list can be regarded as a highly compressed format of utility-list [M. Liu et al. 2012], which is a very efficient structure for high utility itemset mining proposed in recent years. Therefore, PUN-list is much more efficient than utility-list when used to mine high utility itemsets.

(2) An efficient algorithm, called MIP, is developed. After constructing the prefix utility tree (abbreviated as PU-tree) from a database, MIP constructs the PUN-lists of long itemsets by joining the PUN-lists of short itemsets. Meanwhile, MIP directly obtains the utility of an itemset by scanning its PUN-list. To reduce the search space, MIP employs utility upper bound information stored in the PUN-list of an itemset to filter all unpromising itemsets that are descendants of the itemset in a set-enumeration tree.

(3) Extensive experiments on various synthetic and real datasets were conducted to compare MIP with three state-of-the-art algorithms. Experimental results show that MIP substantially outperforms these algorithms. Specially, MIP is on average about an order of magnitude faster than HUI-Miner [M. Liu et al. 2012], which is the fastest one among all recently reported algorithms for high utility itemset mining so far. Our experiments confirm that PUN-list is a very effective data structure for mining high utility itemsets.

The rest of this paper is organized as follows. Section 2 presents the background and related work for high utility itemset mining. Section 3 introduces the PUN-list structure. Section 4 develops a PUN-list-based algorithm, MIP, for high utility item-set mining. Experiment results are shown in Section 5 and conclusions and future work are given in Section 6.

## 2. BACKGROUND

In this section, we first present the formal description of high utility itemset mining and then introduce the related work.

### 2.1 Problem Statement

In this paper, we adopt similar definitions to those presented in previous works [M. Liu et al. 2012; Tseng et al. 2010]. Let $I = \{i_1, i_2, \ldots, i_m\}$ be the universal set of items, $DB = \{T_1, T_2, \ldots, T_n\}$ be a transaction database, and $UT$ be a utility table. Each $T$ in $DB$ is assigned a unique identifier ($Tid$) and is a subset of $I$. In addition, each item in $T$ is associated with a non-zero count value. Independent of transactions, each $i$ in $I$ has an external utility value given by $UT$. We call $A$ an itemset if it is a subset of $I$. $A$ is also called a $k$-itemset if it contains $k$ items.

**Definition 1.** The utility of item $i$ in transaction $T$, denoted as $u(i, T)$, is the product of $c(i, T)$ and $v(i)$, where $c(i, T)$ is the count value associated with $i$ in $T$ and $v(i)$ is the external utility value of $i$ in utility table $UT$.

**Definition 2.** The utility of itemset $A$ in transaction $T$, denoted as $u(A, T)$, is the sum of utilities of all items belonging to $A$ in $T$ which contains $A$, i.e.,

$$u(A,T) = \sum_{i \in A \wedge A \subseteq T} u(i,T).$$

**Definition 3.** The utility of itemset $A$, denoted as $u(A)$, is the sum of utilities of $A$ in all transactions containing $A$ in $DB$, i.e.,

$$u(A) = \sum_{A \subseteq T} u(A,T).$$

**Definition 4.** The utility of transaction $T$, denoted as $tu(T)$, is the sum of the utilities of all items in $T$, i.e.,

$$tu(T) = \sum_{i \in T} u(i,T).$$

**Definition 5.** Let $DB$ be a transaction database, $UT$ a utility table, and $\xi$ a percentage threshold, itemset $A$ is called a high utility itemset if and only if $u(A)$ is not less than $\xi \times \Sigma_{T \in DB} tu(T)$, which is also denoted as *minutility*.

Given a transaction database, a utility table and a threshold $\xi$, the problem of mining high utility itemsets is to discover the set of all itemsets whose utilities are not less than *minutility*.

Table 1: Transaction Database DB_sample

| Tid | Transaction with counts of items |
|---|---|
| $T_1$ | (a, 1), (c, 1), (d, 1), (f, 3) |
| $T_2$ | (a, 2), (b, 2), (c, 6), (f, 5) |
| $T_3$ | (b, 2), (f,5), (g,5) |
| $T_4$ | (b,4), (c,3), (e,2) |
| $T_5$ | (a, 2), (c,2), (d,6), (e,1), (f,1) |

Table 2: Utility Table UT_sample

| Item | a | b | c | d | e | f | g |
|---|---|---|---|---|---|---|---|
| Utility | 30 | 50 | 40 | 30 | 10 | 10 | 20 |

**Example 1**. Let $\{a, b, c, d, e, f, g\}$ be the universal item set, *DB_sample* shown in Table 1 the transaction database, *UT_sample* shown in Table 2 the utility table and *minutility* = 500. In Table 1, each item in a transaction is associated with a non-zero count value. For example, (*a*, 2) in the second transaction ($T_2$) means item *a* occurs twice in the transaction.

Based on Table 1 and Table 2, we know $c(a, T_1) = 1$ and $v(a) = 30$. According to Definition 1, we have $u(a, T_1) = 1 \times 30 = 30$. Similarly, we have $u(a, T_2) = 2 \times 30 = 60$, $u(a, T_5) = 2 \times 30 = 60$, $u(c, T_1) = 1 \times 40 = 40$, $u(c, T_2) = 6 \times 40 = 240$, and $u(c, T_5) = 2 \times 40 = 80$. Therefore, we have $u(\{ac\}, T_1) = u(a, T_1) + u(c, T_1) = 30+40 = 70$, $u(\{ac\}, T_2) = u(a, T_2) + u(c, T_2) = 60+240 = 300$, and $u(\{ac\}, T_5) = u(a, T_5) + u(c, T_5) = 60 + 80 = 140$ by Definition 2. Only $T_1$, $T_2$, and $T_5$ contain {*ac*} in *DB_sample*. Therefore, we have $u(\{ac\}) = u(\{ac\}, T_1) + u(\{ac\}, T_2) + u(\{ac\}, T_5) = 70 + 300 + 140 = 510$ by Definition 3. According to Definition 5, {*ac*} is a high utility itemset as $u(\{ac\}) = 510$, which is greater than 500, the given *minutility*.

### 2.2 Related Work

High utility itemset mining was first introduced in [Chan et al. 2003]. Subsequently, Yao et al. [Yao et al. 2004] formalized the problem of high utility mining and proposed a theoretical model. The main challenge of this mining task is that high utility itemsets do not have the downward closure property. To overcome the challenge, many studies have been proposed, including Two-Phase [Y. Liu et al. 2005], IIDS [Y. C. Li et al. 2008], IHUP [Ahmed et al. 2009], HUC-Prune [Ahmed et al. 2011], UP-Growth [Tseng et al. 2010], and UP-Growth+ [Tseng et al. 2013]. Most of these algorithms work in candidate-generation framework and employ the downward closure property of TWU (transaction-weighted utility) to generate candidates. Two-Phase adapts level-wise, multi-pass candidate generation process like Apriori [Agrawal et al. 1994] to generate candidates called HTWUIs in the first phase. IIDS improves the performance of Two-Phase by discarding isolated items to reduce the number of candidates. IHUP, HUC-Prune, UP-Growth and UP-Growth+ all employ the framework of FP-growth [Han et al. 2000] to generate candidates. These algorithms are distinct in that they employ different tree structures and pruning strategies to discover candidates. Experiments show that these algorithms based on FP-growth generate much fewer candidates than Apriori-based algorithms in the first phase.

Compared with the number of final high utility itemsets, these FP-growth-based algorithms still generate a large number of candidates, and it is very costly to both generate these candidates and compute their exact utilities. To address this issue, HUI-Miner [M. Liu et al. 2012] and d2HUP [J. Liu et al. 2012] are proposed. Both algorithms work in a new framework, which discovers high utility itemsets without generating candidates. In the new framework, high utility itemsets are directly identified from a set-enumeration tree, which is the search space constructed by enumerating itemsets with prefix or suffix extensions. To facilitate the computation of itemsets' utilities and prune unpromising itemsets efficiently, both HUI-Miner and d2HUP maintain utility information and utility upper bound information by two kinds of data structure, utility-list [M. Liu et al. 2012] and CAUL [J. Liu et al. 2012], respectively. By avoiding the generation of a large number of candidates, HUI-Miner and d2HUP show excellent performance and outperform the state-of-the-art algorithms based on FP-growth over one order of magnitude.

However, because utility-list and CAUL are linear with the number of transactions containing an itemset, HUI-Miner and d2HUP will become ineffective when the database becomes dense or large. To solve this problem, we introduce a compressed structure, called prefix utility tree, to store the database and divide the transactions containing an itemset into a collection of mutually dis-joint subsets by storing the transactions in its different nodes. By means of the nodes of the prefix utility tree, we propose a novel structure, PUN-list, to store sufficient utility-relevant information about itemsets. The utility and utility upper bound of an itemset can be computed linearly in terms of the size of its PUN-list. Moreover, the PUN-lists of long itemsets can be constructed by joining the PUN-lists of short itemsets and the computational cost is linear with the size of the PUN-lists of short itemsets. Note that, the size of PUN-lists depends on the compressed prefix utility tree instead of the original database. Therefore, our proposed algorithm, MIP, is insensitive to the size of databases because it employs PUN-list for high utility itemset mining.

Previously, Deng et al. have proposed some kinds of data structure similar to PUN-list, named Node-list [Deng et al. 2010], N-list [Deng et al. 2012], and Nodeset [Deng et al. 2014], to promote the efficiency of frequent itemset mining. However, these kinds of structure store only the summary information about itemsets' counts. Without information about utilities and utility upper bounds, it is not possible to simply adapt them to mine high utility itemsets.

### 3. PUN-LIST STRUCTURE

In this section, we first introduce the procedure of database preprocessing. Then, we present prefix utility tree (PU-tree), which is the basis of PUN-list. Based on PU-tree, the definition and construction of PUN-list are de-scribed in the end.

### 3.1 Database Preprocessing

**Definition 6.** The transaction-weighted utility of itemset $A$ in $DB$, denoted as $twu(A)$, is the sum of the utilities of all the transactions containing $A$ in $DB$, i.e.,

$$twu(A) = \sum\nolimits_{A \subseteq T \wedge T \in DB} tu(T).$$

**Property 1** [Y. Liu et al. 2005]. If $twu(A)$ is less than the given $minutility$, $A$ and all supersets of $A$ are not high utility itemsets.

Property 1 indicates that if the transaction-weighted utility of an item is less than a given $minutility$, the item can be deleted from the database without affecting high utility itemset mining. This provides a way to facilitate the mining process.

**Definition 7.** A succinct database is a database defined below.
(1) The transaction-weighted utility of each item is no less than a given minutility.
(2) Each transaction is associated with its utility.
(3) In each transaction, an item is associated with its utility in the transaction.
(4) Items in each transaction are sorted in item-support descending order.
*(5)* Transactions are sorted in item-support descending order.

Constraint (1) is used to filter unpromising items. Constraint (2) and (3) keep all necessary information for high utility mining. Constraint (4) helps to construct a compressed prefix utility tree. Constraint (5) facilitates the construction of the PUN-lists of 2-itemsets. Note that, item-support descending order is the descending order according to the supports of items in the database.

Previous work [Ahmed et al. 2009; Tseng et al. 2010] suggests that TWU (abbreviation for transaction-weighted utility of item) descending order is suitable for high utility mining. However, our experiments show that item-support descending order is as good as TWU descending order. For the sake of discussion, item-support descending order is used in the remainder of this paper unless otherwise stated.

Table 3 shows the succinct database generated from *DB_sample*. Note that, trans-

6action $T_1$, $T_2$, $T_3$, $T_4$, and $T_5$ in Table 3 correspond to transaction $T_1$, $T_2$, $T_5$, $T_4$, and $T_3$ in Table 1 respectively. Without loss of generality, we still denote $k$th transaction of the succinct database by $T_k$. According to item-support descending order, which is $\{c, f, a, b, e\}$, $\{cfae\}$ is in the third place among all five itemsets corresponding to the five transaction. Therefore, $\{(c, 80), (f, 10), (a, 60), (e, 10)\}$ is the third transaction, denoted as $T_3$, of the succinct database.

Table 3: The succinct version of DB_sample

| Tid | Transaction with item utility | tu |
|---|---|---|
| $T_1$ | (c, 40), (f, 30), (a, 30) | 100 |
| $T_2$ | (c, 240), (f, 50), (a, 60), (b, 100) | 450 |
| $T_3$ | (c, 80), (f, 10), (a, 60), (e, 10) | 160 |
| $T_4$ | (c, 120), (b, 200), (e, 20) | 340 |
| $T_5$ | (f, 50), (b, 100) | 150 |

The last column of Table 3 is used to store the utilities of modified transactions. In addition, each item in a transaction is associated with the item utility in the transaction. For example, $(a, 60)$ in the third row of Table 3 means that the utility of item $a$ in $T_2$ is 60. In Table 3, only $T_1$, $T_2$, and $T_3$ contain item $a$. Therefore, $twu(a)$ is 710, the sum of $tu(T_1)$ (100), $tu(T_2)$ (450), and $tu(T_3)$ (160). Similarly, $twu(b)$, $twu(c)$, $twu(e)$, and $twu(f)$ are 940, 1050, 500, and 860 respectively, all of which are no less than 500, the given *minutility*.

In the remainder of this paper, the succinct database is used to mine high utility itemsets instead of the original database. ***From here on, a database is referred to a succinct database when it is mentioned.***

### 3.2 Prefix Utility Tree

**Definition 8.** Let $I_{SDO}$ be the ordered set by sorting universal item set $I$ in item-support descending order. For item $x$ and $y$, we call $x \succ y$ if $x$ is before $y$ in $I_{SDO}$.

For example, we have $c \succ a$. From here on, itemset $P$ in this paper is denoted as $i_k \ldots i_2 i_1$, where $i_k \succ i_{k-1} \succ \ldots i_2 \succ i_1$, when it is mentioned.

**Definition 9.** Given itemset $A$ ($\subseteq$ transaction $T$), the set of items prior to the first item of $A$ in $T$ are denoted as $prii(A, T)$, i.e.,
$$prii(A, T) = \{i \mid i \in T \wedge i \succ A.first\_item\}.$$

**Definition 10.** Given itemset $A$ ($\subseteq$ transaction $T$), the anterior utility of itemset $A$ in transaction $T$, denoted as $au(A, T)$, is the sum of the utilities of all items of $prii(A, T)$ in $T$, i.e.,
$$au(A,T) = \sum\nolimits_{i \in prii(A,T)} u(i,T).$$

For example, consider itemset $\{ab\}$ and transaction $T_2$ in Table 3, $prii(\{ab\}, T_2) = \{c, f\}$ and $au(\{ab\}, T_2) = 240 + 50 = 290$, where 240 is the utility of item c in $T_2$ and 50 is the utility of item $f$ in $T_2$.

**Definition 11.** Given itemset $A$, the anterior utility of $A$, denoted as $au(A)$, is the sum of the anterior utilities of $A$ in all transactions that contain it, i.e.,
$$au(A) = \sum\nolimits_{A \subseteq T} au(A,T).$$

Before defining the PUN-list structure, we first introduce prefix utility tree, which is the basis of PUN-list. The prefix utility tree is a compact structure that maintains sufficient utility information about items in a database.

**Definition 12.** A prefix utility tree (PU-tree) is a prefix tree defined below.

(1) It is made up of one root labeled as "Root", a set of item prefix subtrees as the children of the root.
(2) Each node consists of five fields: *label*, *n_code*, *Tr_list*, *parent-link*, and *nn-link*. The description of each field is listed as follows.
   (2.1) Field *label* registers which item this node represents.
   (2.2) Field *n_code*, registers a unique identifier which represents the node. In this paper, we use the sequence number of a node by a pre-order traversal of the tree to identify the node.
   (2.3) Field *Tr_list* registers the relevant information about these transactions which register the item in the node. The *Tr_list* of a node consists of a set of triples. Each triple includes three fields: *r_tid*, *r_u*, and *r_au*. *r_tid* registers the Tid of a transaction, *r_u* registers the utility of the item represented by the node in the transaction, and *r_au* registers anterior utility of the item in the transaction.
   (2.4) Field *parent-link* registers the parent of the node.
   (2.5) Field *nn-link* links to the next node which has the same *label*.
(3) Header table is employed to facilitate the traversal of the PU-tree. Each entry in the header table is composed of two fields: *item-name* and *entry-link*. Field *item-name* indicates the item which the entry represents while field *entry-link* points to the first node in the PU-tree whose label is equal to *item-name*.

---

**Algorithm 1:** PU-tree Construction

**Input:** A succinct database *DB*.
**Output:** *PUtr*, a PU-tree.
1: initialize *PUtr* with root *Root*;
2: **For** each transaction *T* in DB **do**
3:    $i \leftarrow T.first\text{-}item$;
4:    $au\_current \leftarrow 0$;
5:    $T.remains \leftarrow T - \{i\}$;
6:    **Call** *Insert_Tree*(*i*, *T.remains*, *Root*);
7: generate the *n_code* of each node by a pre-order traversal of *PUtr*;
8: **Return** *PUtr*;

---

**Function** Insert_Tree(*i*, *T.remains*, *Nd*)

1: Creat *Elem* with *null* as initial value;
2: $Elem.r\_id \leftarrow T.Tid$;
3: $Elem.r\_u \leftarrow u(i,T)$;
4: $Elem.r\_au \leftarrow au\_current$;
5: **If** *Nd* has a child *N* such that *N.label* = *i* **then**
6:    append *Elem* to *N.Tr_list*;
7: **else**
8:    create a new node *N* with item *i* as its *label*;
9:    $N.parent\text{-}link \leftarrow Nd$;
10:   *N.nn-link* is linked to the nodes with item *i* as *label*;
11:   append *Elem* to *N.Tr_list*;
12: **If** $T.remains \neq \varnothing$ **then**
13:   $au\_current \leftarrow au\_current + au(\{i\},T)$;
14:   $i_{next} \leftarrow$ the next item of *i* in *T*;
15:   $T.remains \leftarrow T.remains - \{i_{next}\}$;
16:   **Call** Insert_Tree($i_{next}$, *T.remains*, *N*);

A PU-tree can be constructed with only one scan of the succinct database. Algo-rithm 1 shows the details. Initially, a tree with root *Root* is first created as shown by Line 1. Subsequently, Line 2 to 6 construct the original tree by processing transactions one by one. Each transaction is inserted into the original tree by calling **Insert_Tree**(*i*, *T*, *Nd*). The function **Insert_Tree**(*i*, *T*, *Nd*) is performed as follows. If *Nd* has a child *N* such that *N.label* = *i*, then insert triple (*T.Tid*: $u(i,T)$, $au(i, T)$) into *N.Tr_list* (Note that, $au(i, T)$ is stored in variable *au_current*); else create a new node *N* labeled with *i*, its *parent-link* is linked to *Nd*, its *nn-link* is linked to the nodes with the same *label*, and insert triple (*T.Tid*: $u(i,T)$, $au(i, T)$) into *N.Tr_list*. If the unprocessed part of *T* (denoted as *T.remains*) is nonempty, call **Insert_Tree**($i_{next}$, *T.remains*, *N*) recursively, where $i_{next}$ is the next item to item *i* in *T*. After inserting all transactions, the original tree is constructed. After traversing the original tree in the pre-order to get the *n_code* of each node (Line 7), the construction of the PU-tree is finished and outputted (Line 8).

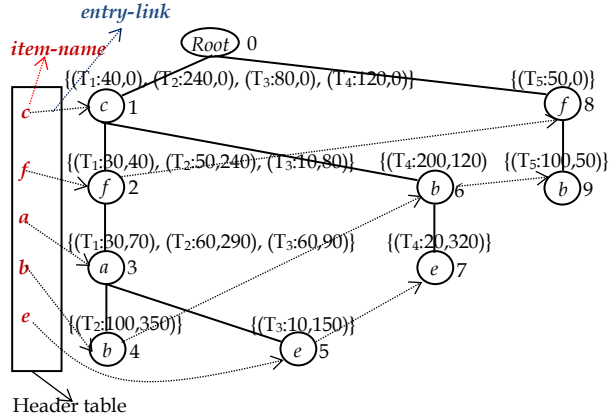

Figure 1: The PU-tree with *minutility* = 500

Figure 1 shows the PU-tree of the database in Table 3 when *minutility* is 500. Let's see the node with label *f* in the left of Figure 1. Number 2 associated with the node is its *n_code*. {($T_1$: 30, 40), ($T_2$: 50, 240), ($T_3$: 10, 80)} associated with the node is its *Tr_list*. The first element, ($T_1$: 30, 40), means that $T_1$ registers utility information about item *f* on the node, and $u(f, T_1)$ and $au(f, T_1)$ are 30 and 40 respectively. For any PU-tree, we have three properties as follows:

**Property 2.** For any node, the triples of its *Tr_list* are sorted in *Tid* ascending order.

**Rationale.** The property can be directly inferred from the first-come-first-served strategy employed by Algorithm 1. □

For example, {($T_1$: 30, 40), ($T_2$: 50, 240), ($T_3$: 10, 80)}, the *Tr_list* of the node with number 2, is sorted in Tid ascending order.

**Property 3.** The nodes following an entry-link are sorted by *n_code* ascending order.

**Rationale.** The property holds since the transactions of the succinct database are sorted by item-support descending order and Algorithm 1 process transactions using the first-come-first-served strategy. □

For example, the nodes following the entry-link of item b are nodes with number 4,

6, and 9, which are sorted in the *n_code* ascending order since 4 < 6 < 9.

**Property 4.** Let $N_1$ and $N_2$ be two nodes registering the same item. If the *n_code* of $N_1$ is bigger than the *n_code* of $N_2$, each *r_tid* in the *Tr_list* of $N_1$ is bigger than any *r_tid* in the *Tr_list* of $N_2$.

**Rationale.** The reason is the same as for Property 3. □

For example, both the node with number 2 and the node with number 8 register item *f*. the *Tr_list* of the former is {($T_1$: 30, 40), ($T_2$: 50, 240), ($T_3$: 10, 80)} while the *Tr_list* of the latter is {($T_5$: 50, 0)}. The *Tid* of $T_5$ is bigger than that of $T_1$, $T_2$, and $T_3$. Note that, the above properties play a key role in efficiently constructing the PUN-lists of 2-itemsets from the PU-tree.

### 3.3 PUN-list: Definition and Construction

Based on the PU-tree, we introduce the structure of PUN-list in this subsection. PUN-list maintains the utility and utility upper bound information about an itemset by summarizing this information at different nodes of a PU-tree. First, we define the PUN-list of 2-itemset. Then, we construct the PUN-list of *k*-itemset by using the PUN-list of (*k*-1)-itemset.

In this paper, the PUN-list of an itemset is an ordered list of quadruples, where each quadruple has four fields: *Nd_id*, *Nd_u*, *Nd_au* and *Nd_aux*. Field *Nd_id* stores a node identifier. We use *n_code* as node identifier. Field *Nd_u* and *Nd_au* record local utility and anterior utility information on a node identified by *Nd_id*. Field *Nd_aux* stores information which facilitates to compute *Nd_u* of *k*-itemset's PUN-list when we join two (*k*-1)-itemset's PUN-lists to construct *k*-itemset's PUN-list.

#### 3.3.1 PUN-lists of 2-itemsets

**Definition 13**. (**PUN-lists of 2-itemsets**) Given a PU-tree, PUtr, and item $i_1$ and $i_2$ ($i_2 \succ i_1$), we denote the set of nodes labeled $i_1$ as $NS_1$, and the set of nodes labeled $i_2$ as $NS_2$. The PUN-list of 2-itemset $\{i_2 i_1\}$, denoted as PUN-list$_{i_2 i_1}$, is constructed as follows: $\forall\ N \in NS_1$, if $\exists\ N' \in NS_2$, $N'$ is an ancestor of $N$, then create a quadruple, (*Nd_id*, *Nd_u*, *Nd_au*, *Nd_aux*), and append it to PUN-list$_{i_2 i_1}$. Concretely, *Nd_id*, *Nd_u*, *Nd_au*, and *Nd_aux* are computed as follows.

(1) *Nd_id* is equal to *N.n_code*, i.e., *Nd_id = N.n_code*. *Nd_id* is used to identify node *N*.

(2) *Nd_u* is the sum of *r_u* of all triples in *N.Tr_list* plus *r_u* of all triples in *N′.Tr_list* with the same *r_tid*, i.e.,

$$Nd\_u = \sum (N.Tr\_list[x].r\_u + N'.Tr\_list[y].r\_u),$$

Where ***N.Tr_list[x].r_tid = N′.Tr_list[y].r_tid.*** Note that *N.Tr_list[x]* represents the *x*th triple of *N.Tr_list* and *N′.Tr_list[y]* represents the *y*th triple of *N′.Tr_list*. *Nd_u* stores the sum of the utilities of {$i_2 i_1$} in all transactions which register item $i_1$ on node *N*.

(3) *Nd_au* is the sum of *r_au* of all triples in *N′.Tr_list* whose *r_tid* occurs in one triple of *N.Tr_list*, i.e.,

$$Nd\_au = \sum N'.Tr\_list[y].r\_au,$$

Where for each *y*, $\exists\ x$ ***N.Tr_list[x].r_tid = N′.Tr_list[y].r_tid.*** *Nd_au* stores the sum of anterior utilities of {$i_2 i_1$} in all transactions which register item $i_1$ on node *N*.

(4) *Nd_aux* is the sum of *r_u* of all triples in *N.Tr_list*, i.e.,

$$Nd\_aux = \sum N.Tr\_list[x].r\_u.$$

*Nd_aux* stores the sum of the utilities of $\{i_1\}$ in all transactions which register item $i_1$ on node $N$. It helps to compute the value of *Nd_u* in PUN-lists of 3-itemsets.

***Note that, for the sake of efficiency, the quadruples in PUN-list$_{i_2i_1}$ are sorted in n_code ascending order.***

According to Definition 13, one key issue of constructing the PUN-list of a 2-itemset is to fast judge the ancestor-descendant relationship of two nodes. Fortunately, the PU-tree with field *parent-link* provides an easy and efficient way to tackle this issue. We will present the method for constructing PUN-lists of 2-itemsets via scanning a PU-tree in Section 4. Here, we introduce our idea of how to construct PUN-lists of 2-itemsets with an example.

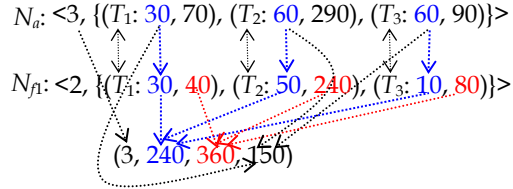

Figure 2: The construction of the PUN-list of $\{fa\}$

Let's examine how to construct the PUN-list of $\{fa\}$, denoted as PUN-list$_{fa}$. According to the PU-tree shown in Figure 1, only one node, whose *n_code* is 3, registers item $a$, and two nodes, whose *n_codes* are 2 and 8 respectively, register item $f$. For the sake of discussion, we denote the only node registering item $a$ as $N_a$, the nodes registering item $f$ as $N_{f1}$ and $N_{f2}$ respectively. According to Figure 1, $N_{f1}$ is the parent of $N_a$. Therefore, a quadruple, denoted as (*Nd_id*, *Nd_u*, *Nd_au*, *Nd_aux*), is constructed and inserted into *PUN-list$_{fa}$*. According to Definition 13, *Nd_id* is equal to 3 ($N_a.n\_code$). From Figure 1, we know that the *Tr_lists* of $N_a$ and $N_{f1}$ are $\{(T_1: 30, 70), (T_2: 60, 290), (T_3: 60, 90)\}$ and $\{(T_1: 30, 40), (T_2: 50, 240), (T_3: 10, 80)\}$ respectively. Because $T_1$, $T_2$, and $T_3$ occur on both of the *Tr_lists*, *Nd_u* is equal to (30+30) + (60 + 50) + (60 + 10) = 240, *Nd_au* is equal to 40 + 240 + 80 = 360, and *Nd_aux* is equal to 30 + 60 + 60 = 150. Figure 2 logically illustrates the processing. According to Figure 1, $N_{f2}$ is not an ancestor of $N_a$. Therefore, the construction is finished. Finally, *PUN-list$_{fa}$* is $\{(3, 240, 360, 150)\}$.

According to the definition of the utility-list of a 2-itemset in [M. Liu et al. 2012], the utility-list of $\{fa\}$ is $\{(T_1: 60, 40), (T_2: 110, 240), (T_3: 70, 80)\}$, where the first number and the second number in each element are the utility and anterior utility of $\{fa\}$ in the corresponding transaction respectively. Clearly, the length (or size) of $\{fa\}$'s utility-list is larger than that of *PUN-list$_{fa}$*, which is $\{(3, 240, 360, 150)\}$. In fact, *PUN-list$_{fa}$* can be regards as the compressed format of $\{fa\}$'s utility-list since it summarizes the utility-relevant information in $\{fa\}$'s utility-list on the node with *n_code* = 3.

The PUN-list of a 2-itemset have two important properties as follows.

**Property 5.** Given 2-itemset $i_2i_1$, the utility of $i_2i_1$ is equal to the sum of the value of all *Nd_u*s in PUN-list$_{i_2i_1}$, i.e.,

$$u(\{i_2i_1\}) = \sum_j PUN-list_{i_2i_1}[j]Nd\_u,$$

Where PUN-list$_{i_2i_1}$ [$j$] represents $j$th quadruple of PUN-list$_{i_2i_1}$.

**Proof:** According to Definition 2 and 3, we have

$$\begin{aligned}u(\{i_2i_1\}) &= \sum_{\{i_2i_1\}\subseteq T} u(\{i_2i_1\},T)\\ &= \sum_{\{i_2i_1\}\subseteq T}(u(i_1,T)+u(i_2,T))\\ &= \sum_{\{i_2i_1\}\subseteq T}u(i_1,T)+\sum_{\{i_2i_1\}\subseteq T}u(i_2,T).\end{aligned} \quad (3.1)$$

Let $ST_{12}$ be the set of all transactions containing both item $i_1$ and $i_2$. According to the construction of a PU-tree, for any $T \in ST_{12}$, it must register item $i_1$ on a node, denoted as $N_{T1}$, and register item $i_2$ on a node, denoted as $N_{T2}$. Furthermore, $N_{T2}$ must be an ancestor of $N_{T1}$. Assume $N_{T1}.Tr\_list[x_T]$ and $N_{T2}.Tr\_list[y_T]$ registers all relevant information of $i_1$ and $i_2$ in transaction $T$ respectively. That is, $N_{T1}.Tr\_list[x_T].r\_u = u(i_1,T)$, $N_{T2}.Tr\_list[y_T].r\_u = u(i_2,T)$, and $N_{T1}.Tr\_list[x_T].r\_tid = N_{T2}.Tr\_list[y_T].r\_tid = T.Tid$, where $T.Tid$ means the Tid of $T$. Therefore,

$$\begin{aligned}\sum_{\{i_2i_1\}\subseteq T}u(i_1,T) &= \sum_{T\in ST_{12}} u(i_1,T)\\ &= \sum_{T\in ST_{12}} N_{T1}.Tr\_list[x_T].r\_u.\end{aligned} \quad (3.2)$$

Similarly, we have

$$\sum_{\{i_2i_1\}\subseteq T}u(i_2,T) = \sum_{T\in ST_{12}} N_{T2}.Tr\_list[y_T].r\_u. \quad (3.3)$$

Plugging Formula (3.2) and Formula (3.3) into Formula (3.1), we have

$$\begin{aligned}u(\{i_2i_1\}) &= \sum_{T\in ST_{12}} N_{T1}.Tr\_list[x_T].r\_u + \sum_{T\in ST_{12}} N_{T2}.Tr\_list[y_T].r\_u\\ &= \sum_{T\in ST_{12}}(N_{T1}.Tr\_list[x_T].r\_u + N_{T2}.Tr\_list[y_T].r\_u).\end{aligned} \quad (3.4)$$

Let $PUNN$ be the set of all nodes of the PU-tree. We denote $\{N \mid (N \in PUNN) \wedge (N.label = i_1) \wedge (\exists\, N' \in PUNN, N'.label = i_2 \text{ and } N' \text{ is an ancestor of } N)\}$ as $NS$. For any $N (\in PUNN)$, we denote the set of all transactions which register item information on $N$ as $TS_N$. According to the construction of PU-tree (Algorithm 1), we have

$$ST_{12} = \bigcup_{N\in NS} TS_N, \ TS_N \cap TS_M = \phi\,(N \neq M).$$

Based on the above conclusion, we can rewrite the right of Formula (3.4) as

$$\begin{aligned}&\sum_{T\in ST_{12}}(N_{T1}.Tr\_list[x_T].r\_u + N_{T2}.Tr\_list[y_T].r\_u)\\ &= \sum_{N\in NS}\sum_{T\in TS_N}(N_j.Tr\_list[v].r\_u + N_j'.Tr\_list[z].r\_u)\end{aligned} \quad (3.5)$$

According to Definition 13, each triple of $PUN\text{-}list_{i2i1}$ is generated by a node in $NS$ and the inner $\Sigma$ is just the $Nd\_u$ of a triple. Therefore, the right of Formula (3.5) is $\Sigma_j PUN\text{-}list_{i2i1}[j].Nd\_u$. Together with Formula (3.4) and Formula (3.5), Property 5 holds. □

**Property 6.** Given 2-itemset $i_2i_1$, the anterior utility of $i_2i_1$ is equal to the sum of all $Nd\_au$s in PUN-list$_{i2i1}$, i.e.,

$$au(\{i_2i_1\}) = \sum_j PUN-list_{i_2i_1}[j].Nd\_au.$$

**Proof:** Let $ST_{12}$ be the set of all transactions containing both item $i_1$ and $i_2$. According to Definition 11, we have

12$$au(\{i_2i_1\}) = \sum_{T \in ST_{12}} au(\{i_2i_1\},T). \tag{3.6}$$

According to Definition 9, $prii(\{i_2i_1\}, T)$ is equal to $prii(\{i_2\}, T)$ Since $i_2 \succ i_1$. Therefore, $au(\{i_2i_1\}, T)$ is equal to $au(\{i_2\}, T)$ according to Definition 10. Thus, we can rewrite Formula (3.6) as

$$au(\{i_2i_1\}) = \sum_{T \in ST_{12}} au(\{i_2\},T). \tag{3.7}$$

By using $NS$ and $TS_N$ defined in the proof of Property 5, we can rewrite Formula (3.7) as

$$au(\{i_2i_1\}) = \sum_{N \in NS} \sum_{T \in TS_N} au(\{i_2\},T). \tag{3.8}$$

Note that, $N'.Tr\_list[y].r\_au$ used to compute $Nd\_au$ registers the anterior utility of item $i_2$ in a transaction of $ST_{12}$. Therefore, the inner $\Sigma$ of Formula (3.8) is just the $Nd\_au$ of a triple. In addition, $PUN\text{-}list_{i_2i_1}$ is generated from the nodes in $NS$ according to Definition 13. Therefore, the right of Formula (3.8) is $\Sigma_j PUN\text{-}list_{i_2i_1}[j].Nd\_au$. Thus, Property 6 holds. □

For better understanding of Property 5 and 6, let's see an example. $Nd\_u$ and $Nd\_au$ of $PUN\text{-}list_{fa}$ are 240 and 360, which are equal to $u(\{fa\})$ and $au(\{fa\})$ respectively.

### 3.3.2 PUN-lists of *k*-itemsets (*k* ≥ 3)

**Definition 14. (PUN-lists of *k*-itemsets)** Let $A(= \{i_k i_{k-1} i_{k-2} \ldots i_2 i_1\})$ be a *k*-itemsets. The PUN-lists of $A_1(= \{i_{k-1} i_{k-2} \ldots i_2 i_1\})$ and $A_2(= \{i_k i_{k-2} \ldots i_2 i_1\})$ are denoted as PUN-list$_{A_1}$ and PUN-list$_{A_2}$ respectively. The PUN-list of $A$, denoted as PUN-list$_A$, is constructed by following procedures: $\forall\ Tp \in$ PUN-list$_{A_1}$, if $\exists\ Tp^* \in$ PUN-list$_{A_2}$ such that $Tp^*.Nd\_id = Tp.Nd\_id$, then create a quadruple, ($Nd\_id$, $Nd\_u$, $Nd\_au$, $Nd\_aux$), and insert it to PUN-list$_A$. $Nd\_id$, $Nd\_u$, $Nd\_au$, and $Nd\_aux$ are computed as follows:

$Nd\_id = Tp.N\_id$;
$Nd\_u = Tp.Nd\_u + Tp^*.Nd\_u - Tp.Nd\_aux$;
$Nd\_au = Tp^*.Nd\_au$;
$Nd\_aux = Tp.Nd\_u$.

***Similar to Definition 13, for the sake of efficiency, the quadruples in PUN-list$_A$ are sorted in n\_code ascending order.***

According to Definition 14, the PUN-list of a *k*-itemset can be constructed by comparing the $Nd\_ids$ in the PUN-lists of two (*k*-1)-itemset using 2-way comparison. Assume the lengths of the PUN-lists of two (*k*-1)-itemset are $m$ and $n$ respectively. It is at most ($m + n$) comparisons for constructing the *k*-itemset's PUN-list since each (*k*-1)-itemset's PUN-list is ordered in $n\_code$ ascending order. For the same reason, the *k*-itemset's PUN-list is naturally ordered in $n\_code$ ascending order.

Let's examine how to construct the PUN-list of {*cfb*}, $PUN\text{-}list_{cfb}$. We know $PUN\text{-}list_{fb}$ and $PUN\text{-}list_{cb}$ are {(4, 150, 240, 100), (9, 150, 0, 100)} and {(4, 340, 0, 100), (6, 320, 0, 200)} respectively. Only 4 is the only value of $Nd\_id$ shared by $PUN\text{-}list_{fb}$ and $PUN\text{-}list_{cb}$. Therefore, $PUN\text{-}list_{cfb}$ contains only one quadruple, whose $Nd\_id$, $Nd\_u$, $Nd\_au$, and $Nd\_aux$ are 4, 390 (150 + 340 – 100), 0, and 150 respectively. Figure 3 illustrates the procedure.

PUN-lists of $k(\geq 3)$-itemsets also have two important properties which are similar to Property 5 and Property 6 respectively. These properties can be established by the inference methods used in the proof of Property 5 and Property 6. Limited by space, we present these properties while omitting their proof.

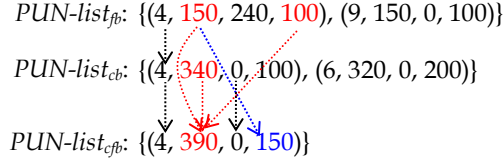

$PUN\text{-}list_{fb}$: {(4, 150, 240, 100), (9, 150, 0, 100)}

$PUN\text{-}list_{cb}$: {(4, 340, 0, 100), (6, 320, 0, 200)}

$PUN\text{-}list_{cfb}$: {(4, 390, 0, 150)}

Figure 3: The construction of the PUN-list of {$cfb$}

**Property 7.** Given $k(≥3)$-itemset $A$, the utility of $A$ is equal to the sum of the value of all $Nd\_u$s in PUN-list$_A$, i.e.,

$$u(A) = \sum_j PUN-list_A[j].Nd\_u.$$

**Property 8.** Given $k(≥3)$-itemset $A$, the anterior utility of $A$ is equal to the sum of the value of all $Nd\_au$s in PUN-list$_A$, i.e.,

$$au(\{A\}) = \sum_j PUN-list_A[j].Nd\_au.$$

## 4 MIP: MINING HIGH UTILITY ITEMSETS USING PUN-LIST

In this section, we introduce our algorithm, MIP, for high utility itemset mining. Besides using PUN-List structure to store sufficient utility information of itemsets, MIP adopts set-enumeration tree [Rymon 1992], which has proven effective in frequent itemset mining [Zaki et al. 2003] and high utility itemset mining [M. Liu et al. 2012; J. Liu et al. 2012], as the search space to facilitate the mining process. In addition, MIP also employs a pruning strategy to reduce the search space.

### 4.1 Search Space and Pruning Strategy

**Definition 15.** Let $I_{SDO}$ be the universal item set, which is sorted in item-support descending order. Without loss of generality, we denote $I_{SDO}$ as {$i_m$, …, $i_2$, $i_1$}. A set-enumeration tree is defined as follows. Firstly, the root of the tree is built; secondly, the $m$ child nodes of the root representing $m$ 1-itemset are built; thirdly, for a node representing itemset {$i_u… i_v$} ($1 ≤ v < u ≤ m$), the $(m - v)$ child nodes of the node representing {$i_{u+1}i_u…i_v$}, {$i_{u+2}i_u…i_v$}, …, {$i_m i_u…i_v$} are sequentially created from left to right.

Consider Example 1. $I_{SDO}$ is {$c$, $f$, $a$, $b$, $e$}, where $c ≻ f ≻ a ≻ b ≻ e$. Figure 4 illustrates the set-enumeration tree of all subsets of $I_{SDO}$.

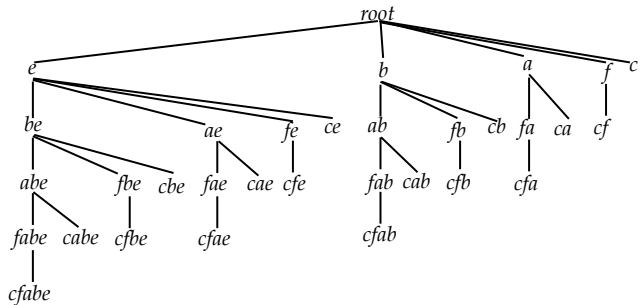

Figure 4: An example of Set-Enumeration Tree

**Definition 16.** Given a set-enumeration tree, the itemset represented by a node is called an ***extension*** of the itemset represented by an ancestor node of the node.

For an itemset containing $k$ items, we call its extension containing with $(k + j)$ items a $j$-extension of the itemset. For example, in Figure 4, itemsets $\{cfae\}$ is a 2-extension of $\{ae\}$.

**Lemma 1.** Given itemset $A$, if $u(A)$ plus $au(A)$ is less than the given *minutility*, any extension of $A$ is not a high utility itemset.

**Proof**. Let $P$ be an extension of $A$. we denote the itemset, generated by deleting all items in $A$ from $P$, as $P/A$. We have

$$\begin{aligned}
u(P) &= \sum_{P \subseteq T} u(P, T) \\
&= \sum_{P \subseteq T} \sum_{i \in P} u(i, T) \\
&= \sum_{P \subseteq T} (\sum_{i \in A} u(i, T) + \sum_{i \in P/A} u(i, T)) \\
&= \sum_{P \subseteq T} \sum_{i \in A} u(i, T) + \sum_{P \subseteq T} \sum_{i \in P/A} u(i, T)).
\end{aligned} \qquad (4.1)$$

It is clear that $A \subseteq P \Rightarrow \{ T \mid P \subseteq T \} \subseteq \{ T \mid A \subseteq T \}$. By replacing $P \subseteq T$ with $A \subseteq T$ in Formula (4.1), we have

$$\begin{aligned}
u(P) &\leq \sum_{A \subseteq T} \sum_{i \in A} u(i, T) + \sum_{A \subseteq T} \sum_{i \in P/A} u(i, T)) \\
&= u(A) + \sum_{A \subseteq T} \sum_{i \in P/A} u(i, T)).
\end{aligned} \qquad (4.2)$$

For any item $i \in (P/A - prii(A, T))$, since $P$ is an extension of $A$ and any itemset is sorted by $\succ$ as mentioned in Subsection 3.2, we have $i \in P/A \Rightarrow \forall\ i_x \in A, i \succ i_x$. Thus, if $i \in T$, we have $i \in prii(A, T)$, which is contradictory to $i \in (P/A - prii(A, T))$. Therefore, for any item $i \in (P/A - prii(A, T))$, we know $i \notin T$. In addition, we know $\forall\ i \notin T, u(i, T) = 0$ according to Definition 1. Based on the above discussions, we have

$$\begin{aligned}
\sum_{A \subseteq T} \sum_{i \in P/A} u(i, T)) &= \sum_{A \subseteq T} \sum_{i \in (P/A) \cap prii(A,T)} u(i, T)) + \sum_{A \subseteq T} \sum_{i \in (P/A) - prii(A,T)} u(i, T) \\
&= \sum_{A \subseteq T} \sum_{i \in (P/A) \cap prii(A,T)} u(i, T)) + \sum_{A \subseteq T} \sum_{i \in (P/A) - prii(A,T)} 0 \\
&= \sum_{A \subseteq T} \sum_{i \in (P/A) \cap prii(A,T)} u(i, T)) \\
&\leq \sum_{A \subseteq T} \sum_{i \in prii(A,T)} u(i, T))
\end{aligned} \qquad (4.3)$$

According to Definition 10 and Definition 11, we know

$$au(A) = \sum_{A \subseteq T} \sum_{i \in prii(A,T)} u(i, T). \qquad (4.4)$$

Plugging Formula (4.4) into Formula (4.3), we have

$$\sum_{A \subseteq T} \sum_{i \in P/A} u(i, T)) \leq au(A). \qquad (4.5)$$

Together with Formula (4.2) and Formula (4.5), we have $u(P) \leq u(A) + au(A)$. Therefore, if $u(A) + au(A) < minutility$, we have $u(P) < minutility$. Lemma 1 holds. □

Note that, Lemma 1 is similar to the principles which were employed by [M. Liu et al. 2012; J. Liu et al. 2012] as the main pruning strategy.

### 4.2 MIP Algorithm

Based on properties in Subsection 3.3 and Lemma 1, we develop an efficient algorithm, named MIP, for mining high utility itemsets using PUN-list. MIP performs a depth-first search of the set-enumeration tree and employs Lemma 1 as the pruning strategy to narrow the search space. Algorithm 2 shows the pseudo-code of MIP.

---
**Algorithm 2:** MIP Algorithm
---
**Input:** *DB*, a transaction database with utility table, and *minutility*, a given threshold.
**Output:** *HUI_set, the set of all high utility itemsets.*
1: scan *DB* to obtain the succinct database, $I_{SDO}$, the item set sorted in support descending order, and each item's utility, and then *HUI_set* ← { $i$ | $x \in I_{SDO} \wedge u(\{i\}) \geq minutility$ };
2: run Algorithm 1 to construct the PU-tree;
3: **For** each item $i$ in the header table of the PU-tree **do**
4:     Assign $Node_{current}$ to the first node labeled with $i$;
5:     $pr_i \leftarrow \varnothing$;
6:     **While** $Node_{current}$ is not null **do**
7:       $N_{ancestor} \leftarrow Node_{current}.parent\text{-}link$;
8:       **While** $N_{ancestor}$ is not node *root* **do**
9:         $x \leftarrow N_{ancestor}.label$;
10:         **If** $x \notin pr_i$, **then** $pr_i \leftarrow pr_i \cup \{ x \}$;
11:         scan the *Tr_list*s of $Node_{current}$ and $N_{ancestor}$, and use 2-way comparison to obtain an element of *PUN-list$_{xi}$*, the *PUN-list* of itemset $xi$;
12:         $N_{ancestor} \leftarrow N_{ancestor}.parent\text{-}link$;
13:       $Node_{current} \leftarrow Node_{current}.nn\text{-}link$;
14:     **For** each $x$ in $pr_i$ **do**
15:       scan *PUN-list$_{xi}$* to obtain $u(xi)$ and $au(xi)$;
16:       **If** $u(xi) \geq minutility$ **then**
17:         *HUI_set* ← *HUI_set* ∪ {$xi$};
18:       **For** each $xi$ **do**
19:         **If** $u(xi) + au(xi) \geq minutility$ **then**
20:         **Call SHUI**( $xi$, {$yi$ | $y \in pr_i \wedge y \succ x$ });
21: **Return** *HUI_set;*

---

The input for MIP is *DB*, a transaction database with utility table, and *minutility*, a given threshold. Line 1 preprocesses the original database to construct the succinct database and thus to obtain relevant initial information. Line 2 employs the PU-tree Construction algorithm (Algorithm 1) to build the PU-tree. The procedure of mining high utility itemset mining consists of Line 3 to 20. For each outmost loop initiated by Line 3, MIP discovers all high utility itemsets that are extensions of item *i*.

Line 4 obtains the first node in the PU-tree whose label is *i*. Line 5 initialize $pr_i$, the set of items that are used to extend item *i* to generate its 1-extensions. Line 6 to 13 generate the PUN-lists of these 2-itemsets, which are *i*'s 1-extensions, by scanning each path from a node labeled with item *i* to node *root*. Line 11 employs Definition 13 to construct the PUN-lists of 2-itemsets. Line 15 obtain the utility and the anterior utility of itemset $xi$, $u(xi)$ and $au(xi)$, by scanning *PUN-list$_{xi}$*. For each 1-extension of item *i*, Line 16 and 17 examine whether it is a high utility itemset. Line 18 and 20 omit unpromising 1-extension by employing Lemma 1. Given a 1-extension, only if the sum of its utility and anterior utility is not less than *minutility*, it can be further used to search high utility itemsets by calling function **SHUI()**. By recursively calling **SHUI()**, MIP finds all high utility itemsets.

Function **SHUI(**$yA$, {$zA$ | $z \succ y$}**)** is used to find all high utility itemsets from the whole set of extensions of itemset $yA$. The process of **SHUI()** is the same as that of the main body (Line 14 to 20) of MIP except for the construction of PUN-list. As mentioned in Subsection 3.3.2, the PUN-list of a $k(\geq 3)$-itemset can be efficiently built from the PUN-lists of two ($k$-1)-itemset by simply using 2-way comparison.

| **Function:** SHUI( $yA$, $\{zA \mid z \succ y\}$ ) |
|---|
| 1: **If** $\{zA \mid z \succ y\} \neq \varnothing$ **then** |
| 2:    **For** each $z \succ y$ **do** |
| 3:      construct $PUN\text{-}list_{zyA}$ via PUN-list$_{zA}$ and PUN-list$_{yA}$; |
| 4:      scan the $PUN\text{-}list_{zyA}$ to obtain $u(zyA)$ and $au(zyA)$; |
| 5:      **if** $u(zyA) \geq minutility$ **then** |
| 6:        $HUI\_set \leftarrow HUI\_set \cup \{zyA\}$; |
| 7:    **For** each $zyA$ **do** |
| 8:      if $u(zyA) + au(zyA) \geq minutility$ **then** |
| 9:        **Call** SHUI($zyA$, $\{vyA \mid v \succ z\}$); |

*It should be noticed that we employ some trick in the implementation of constructing the PUN-lists of 1-extensions (Line 11 in Algorithm 2) for the sake of efficiency.* In fact, we set a pointer *mark* for each $N_{ancestor}$, which records the triple whose $r\_tid$ is equal to the largest $r\_tid$ in the $Tr\_list$ of N, the node pointed by $Node_{current}$. When $Node_{current}$ point to $N_{next}$, the next node to N along N's *nn-link*, we only need to compare $N_{ancestor}$ with $N_{next}$ from the triple next to the triple recorded by *mark* if $N_{ancestor}$ is an ancestor of $N_{next}$. That is, we need not to start from the first triple of $N_{ancestor}$. The reasonability is ensured by Property 2, 3 and 4. According to Property 3, the $n\_code$ of $N_{next}$ is bigger than the $n\_code$ of $N_{next}$. Further, according to Property 4, each $r\_tid$ in the $Tr\_list$ of $N_{next}$ is bigger than any $r\_tid$ in the $Tr\_list$ of N. That is, each $r\_tid$ in the $Tr\_list$ of $N_{next}$ is bigger than the $r\_tid$ of the triple of $N_{ancestor}$, which is recorded by *mark*. Finally, according to Property 2, each $r\_tid$ in the $Tr\_list$ of $N_{next}$ is bigger than the $r\_tid$ of each triple that is ahead of the triple recorded by *mark* in $N_{ancestor}$. Therefore, to find triples in $N_{next}$ and $N_{ancestor}$ with the same $r\_tid$, it can start from the triple next to the triple recorded by *mark*. Of course, *mark* should be reinitialized to null for each outmost loop of Algorithm 2.

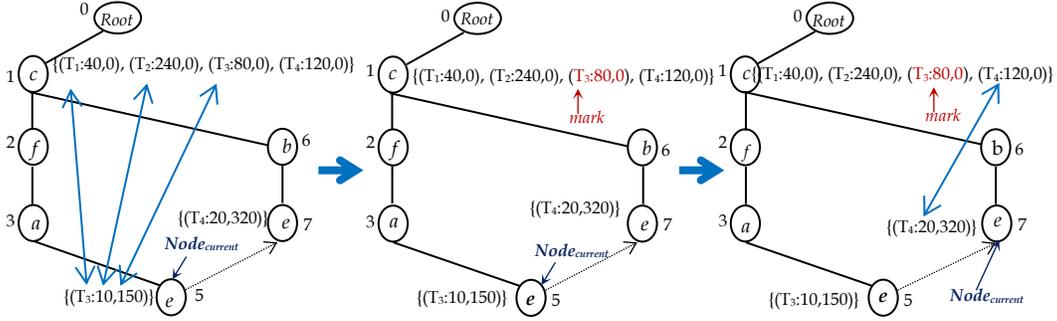

Figure 5: The process of constructing the PUN-lists of itemset {*ce*}

For better understanding of the trick, let's take Figure 1 as an example to examine how to construct PUN-list$_{ce}$, the PUN-list of itemset *ce*. For the sake of discussion, we denote a node by its $n\_code$. First, $Node_{current}$ points to node 5, the first node registering item *e*. Then, along *parent-link*, $N_{ancestor}$ backtracks to node *root* from node 5. When $N_{ancestor}$ reach node 1, $\{(T_3: 10, 150)\}$, the $Tr\_list$ of node 5, and $\{(T_1: 40, 0), (T_2: 240, 0), (T_3: 80, 0), (T_4: 120, 0)\}$, the $Tr\_list$ of node 1, are scanned via 2-way comparison to get quadruple (5, 90, 0, 10). Then, (5, 90, 0, 10) is appended to PUN-list$_{ce}$ and the *mark* of node 1 is set to point to $(T_3: 80, 0)$. When $N_{ancestor}$ reaches node *root*, the process for this path is finished and $Node_{current}$ points to node 7, the next node to node 5 along *nn-link*. Once again, $N_{ancestor}$ backtracks to node *root* from node 7 along *parent-link*. When $N_{ancestor}$ reaches node 1 again, not $\{(T_1: 40, 0), (T_2: 240, 0), (T_3: 80, 0), (T_4: 120, 0)\}$ but $\{(T_4: 120, 0)\}$ is compared with $\{(T_4: 20, 320)\}$, the $Tr\_list$ of node 7, since the *mark* of node 1 points to ($T_3$: 80, 0). After comparison, quadruple (7, 140, 0, 20) is computed and appended to PUN-list$_{ce}$. So far, no node registering item *e* is following node 7 along *nn-link*. The construction of PUN-list$_{ce}$ is finished and it is {(5, 90, 0, 10), (7, 140, 0, 20)}. Figure 5 illustrates the processes. This example shows that our trick greatly avoids lots of useless computation.

## 5 EXPERIMENTAL EVALUATION

In this section, we present a comprehensive performance comparison of MIP with the state-of-the-art mining algorithms on various real and synthetic datasets. We first introduce the experimental setup and then report our experimental results in terms of runtime, memory consumption, and scalability.

### 5.1 Experimental Setup

To evaluate the performance of MIP, we selected several algorithms as compared algorithms. These algorithms are HUI-Miner [M. Liu et al. 2012], d$^2$HUP [J. Liu et al. 2012], and UP-Growth+ [Tseng et al. 2013]. The three state-of-the-art algorithms had been proven to outperform other algorithms for mining high utility itemsets. All algorithms were implemented in C++ language and compiled using g++. Two versions of MIP, denoted as MIP_sup and MIP_twu, were implemented. MIP_sup employs item-support descending order to build the PU-tree while MIP_twu employs TWU descending order to build the PU-tree.

Table 4: Characteristics of datasets

| Dataset | #Trans | #Items | AveLen | MaxLen |
| --- | --- | --- | --- | --- |
| Connect | 67,557 | 129 | 43 | 43 |
| Pumsb | 49,046 | 2,113 | 74 | 74 |
| Accidents | 340,183 | 468 | 33.8 | 51 |
| Chess | 3,196 | 75 | 37 | 37 |
| Mushroom | 8,124 | 119 | 23 | 23 |
| Chain | 1,112,949 | 46,086 | 7.3 | 170 |
| T10I4D100K | 100,000 | 870 | 10.1 | 29 |
| T40I10D100K | 100,000 | 942 | 39.6 | 77 |

The experiments were performed on a 3.20GHz PC machine (Intel Core i5-4570) with 8GB of memory, running on an Ubuntu14.04 operating system. Eight datasets were used in the experiments to evaluate these algorithms extensively. Dataset *connect*, *pumsb*, *accidents*, *chess*, and *mushroom* downloaded from FIMI Repository (http://fimi.ua.ac.be/, 2013) are real datasets. Dataset *chain* was downloaded from NU-MineBench (http://cucis.ece.northwestern.edu/projects/DMS/MineBench.html, 2013). Dataset *T10I4D100K* and *T40I10D100K* are synthetic datasets generated from the data generator in [Agrawal et al. 1994]. Table 4 shows the statistical information about these datasets, including the number of transactions, the number of distinct items, the average number of items in each transaction, and the maximal number of items in the longest transaction(s). We generate external utilities and counts of items according to the methods used in [M. Liu et al. 2012] for these datasets except *chain* which provides such information. External utilities of items are generated between 0.01 and 10 using a log-normal distribution. Counts of items are generated randomly ranging from 1 to 10.

### 5.2 Runtime Consumption

***Note that runtime here means the total execution time, which is the period between input (original databases) and output (all high utility itemsets).*** For

example, the runtime of MIP (both MIP_sup and MIP_twu) includes the time of transforming original databases into succinct databases, constructing PU-tree, and generating the PUN-lists of itemsets.

Like the performance evaluation of previous algorithms, we varied the minimum utility threshold for each dataset. Note that the minimum utility threshold is the percentage of *minutility* over the sum of utilities of all items in the database.

Figure 6 shows the runtime of all algorithms on eight datasets. When the minimum utility threshold decreases, more runtime is needed since more high utility itemsets are found.

In Figure 6, MIP_sup and MIP_twu always run much faster than other three algorithms on all datasets and minimum utility thresholds. The curves for MIP_sup and MIP_twu overlap each other, which means that they have almost the same efficiency. HUI-Miner performs better than $d^2$HUP and UP-Growth+. Although $d^2$HUP outperforms UP-Growth+, they are both at least one order of magnitude slower than MIP_sup, MIP_twu, and HUI-Miner. Specially, UP-Growth+ cannot find all high utility itemsets from some datasets in a reasonable time. For example, UP-Growth+ runs over 10,000 seconds on dataset *accidents* when the minimum utility threshold is less than 13%. On dataset *connect*, *chess*, and *pumsb*, UP-Growth+ runs out of time even under the largest minimum utility threshold. Note that, our experimental results are consistent with those reported in [Lan et al. 2014] except the results on dataset *T40I10D100K*. The reason may be that the data distribution of our *T40I10D100K* is significantly different from that of the one used in [M. Liu et al. 2012] since they were generated randomly according to some probability function.

Generally, MIP_sup and MIP_twu are almost two orders of magnitude faster than HUI-Miner on dataset *connect* and about one orders of magnitude faster than HUI-Miner on other datasets except *T10I4D100K*. Although MIP_sup and MIP_twu run faster than HUI-Miner on *T10I4D100K*, their advantage is not as great as on other datasets. The reason can be described as follows. Since *T10I4D100K* is very sparse, its PU-tree is big and not compressed enough. Therefore, the PUN-lists built from the PU-tree are not short enough. In fact, the advantage of MIP over HUI-Miner mainly lies in that PUN-lists are much shorter and more compressed than utility-lists used in HUI-Miner. However, the length (or compression capability) of PUN-lists depends on datasets. Generally, the denser a dataset is, the more compressed its PU-tree is, and thus the shorter of the PUN-lists. Table 5 shows the average PUN-list (in MIP_twu) and utility-list (in HUI-Miner) lengths across all high utility $k(\geq 2)$-itemsets on all eight datasets. Collating Table 5 with Figure 6, we find that as the reduction ratio increases, MIP performs better. This observation also supports our argument above.

In addition, compared with HUI-Miner, MIP also has an advantage on mining high utility 2-itemsets. Since the PU-tree partition all transaction containing an item into different groups attached to different nodes, MIP can employ divide-and-conquer method to obtain 2-itemsets by computing at group-level. However, HUI-Miner needs to scan all transactions of each items of a 2-itemsets, which is inefficient than MIP. Table 6 shows the time consumption for mining high utility 2-itemsets and all high utility itemsets when HUI-Miner and MIP_twu run on dataset *chain* with minimum utility = 0.005%. This table also explains the phenomenon, where MIP_twu still runs much faster than HUI-Miner on dataset *chain* even though the average PUN-list cardinality is almost the same as the average Utility-list cardinality.

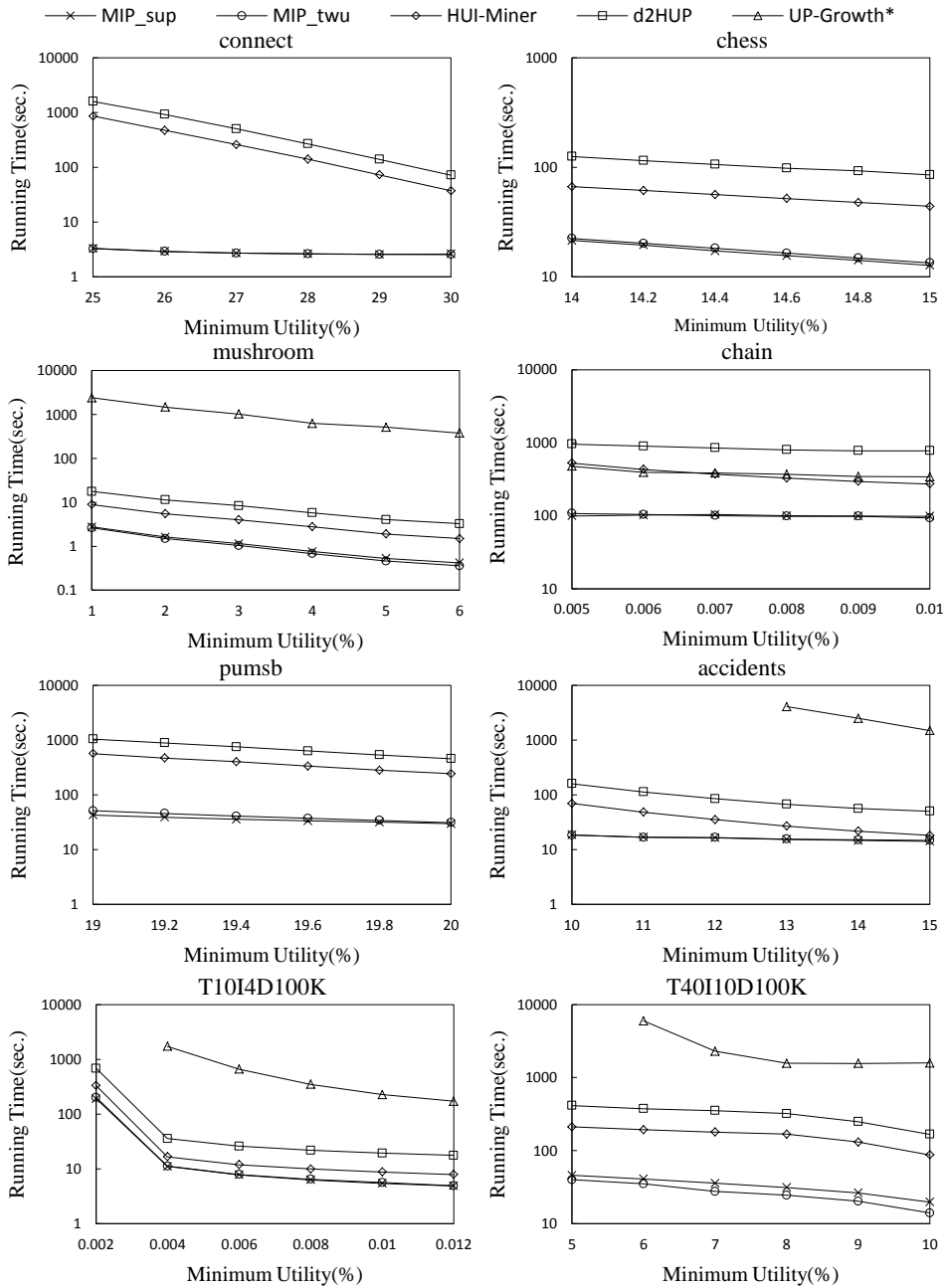

Figure 6: Runtime Comparison

Table 5: Average Utility-list and PUN-list Cardinality
(Reduction Ratio = Avg. Utility-list Length / Avg. PUN-list Length)

| Dataset | $min\_util$(%) | Avg. Utility-list Length | Avg. PUN-list Length | Reduction Ratio |
|---|---|---|---|---|
| Connect | 25 | 42,949.3 | 21 | 2045.2 |
| Pumsb | 19 | 32,004.5 | 26.3 | 1216.9 |
| Accidents | 10 | 141,243.7 | 1,051.3 | 134.4 |
| Chess | 14 | 1,235.9 | 336.3 | 3.7 |

| Mushroom | 2 | 491.4 | 141.9 | 3.5 |
| Chain | 0.005 | 325.3 | 231.2 | 1.4 |
| T10I4D100K | 0.002 | 1.13 | 1.12 | 1 |
| T40I10D100K | 5 | 26,541.5 | 3,367.9 | 7.9 |

Table 6: Runtime for 2-itemsets and all itemsets

| Algorithm | Time(2-itemsets) | Time(all itemsets) |
|---|---|---|
| MIP_twu | 46.93 | 107.48 |
| HUI-Miner | 365.9 | 475.08 |

**5.3 Memory Consumption**

In this subsection, we present and discuss memory consumption of all five algorithms. Figure 7 shows the peak memory consumption of the compared algorithms under varied minimum utility on eight datasets. MIP_sup and MIP_twu consume almost the same memory. Similar to the case occurring in runtime, their curves overlap again. In the figure, HUI-Miner performs best on the whole since it uses the lowest amount of memory on four out of all eight datasets. MIP performs best on dataset *pumsb* and *connect*. Depending on dataset, it consumes 0.4 to 5 times the memory of HUI-Miner. Although UP-Growth+ uses the lowest amount of memory on dataset *accidents* and *T40I10D100K* when the minimum utility threshold is high, it is inefficient since it runs out of time when the minimum utility threshold is low, as shown in Figure 6.

Since MIP and HUI-Miner employ similar structure and they run much faster than $d^2HUP$ [J. Liu et al. 2012] and UP-Growth+, we focus on comparing MIP and HUI-Miner in following discussion.

In Figure 7, we observe that compared with HUI-Miner, MIP performs better on dense datasets. Next, we take dataset *connect* and *T10I4D100K* as an example to discuss the reason since they are typical dense and sparse datasets respectively.

MIP consumes less memory than HUI-Miner on dataset *connect* while it consumes more memory than HUI-Miner on dataset *T10I4D100K*. As we know, the memory consumption depends on two parts, the size of the necessary data structure and that of intermediate results during the mining process. For MIP, the necessary data structure is the PU-tree. For HUI-Miner, the necessary data structure is the utility-lists of items, which is actually the dataset. The PU-tree contains not only the dataset but also nodes and links. Therefore, the PU-tree is larger than the dataset. The size of intermediate results in MIP is the sum of the sizes of their PUN-lists while the size of intermediate results in HUI-Miner is the sum of the sizes of their utility-lists. Since MIP and HUI-Miner employ similar pruning strategy, the number of intermediate results is the same for them. Therefore, the size of intermediate results depends on the average size of PUN-lists or utility-lists.

As mentioned in Subsection 5.2, the average length of PUN-lists is three orders of magnitudes smaller than that of utility-lists on dataset *connect*. This suggests that the size of intermediate results in MIP is much less than that in HUI-Miner on dataset *connect*. In addition, since *connect* is a very dense dataset, its PU-tree is highly compressed in terms of tree structure. Thus, the PU-tree is a little bigger than the dataset. Based on the above analysis, MIP performs better than HUI-Miner on dataset *connect*. For dataset *T10I4D100K*, the situation is just the opposite. First, *T10I4D100K* is a very sparse dataset. Second, the reduction ratio of between average length of PUN-lists and that of utility-lists is not big enough. Therefore, HUI-Miner uses less memory than MIP on *T10I4D100K*.

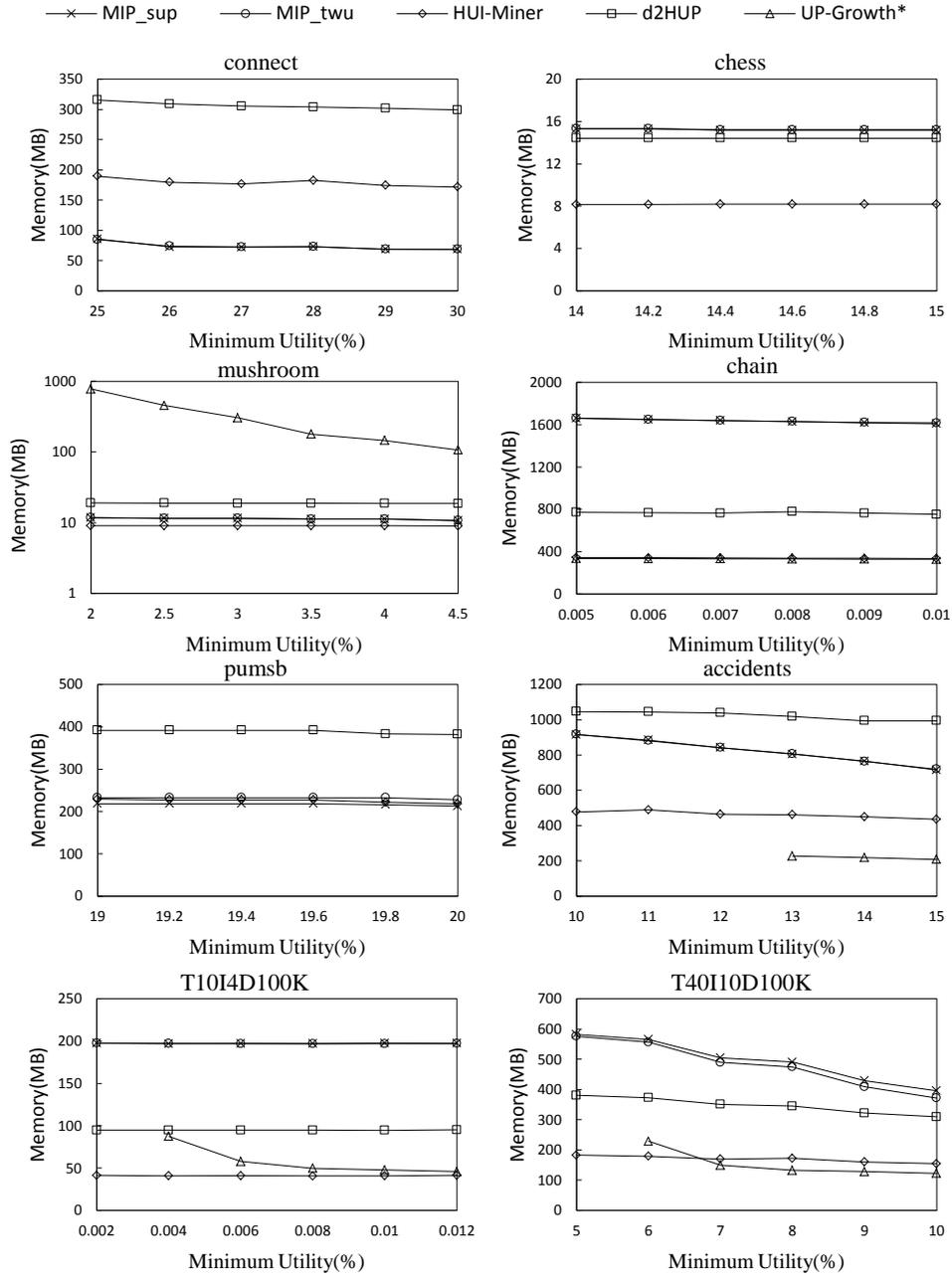

Figure 7: Memory Comparison

Generally, when a dataset is dense and reduction ratio is large, MIP consumes less memory than HUI-Miner and vice versa. However, it is very difficult to quantitatively analyze the memory usage since the data distributions of datasets are various and hard to be explored.

### 5.3 Scalability

In this paper, we only compare the scalability of MIP_sup, MIP_twu, and HUI-Miner since UP-Growth+ and d²HUP run much slower than them in terms of runtime. The scalability test is conducted to evaluate the runtime performance and memory consumption of MIP and HUI-Miner on large-scale datasets.

The experiments are performed on two groups of dataset. The first group, denoted by *G_T40I10*, includes five datasets. These datasets are generated by the data generator in [Agrawal et al. 1994] with the same parameter setting as *T10I4D100K* except that the number of transactions is set to 100K, 200K, 300K, 400K, and 500K respectively. The second group, denoted by *G_accidents*, also includes five datasets, which contains 1M, 2M, 3M, 4M, and 5M transactions respectively. They are generated by randomly selecting transactions from dataset *accidents* via sampling with replacement. For example, to generate the dataset with 1M transactions, we randomly sample 1M transactions from dataset *accidents* with replacement. Note that, all datasets in each group share the same item external utility. For datasets from *G_T40I10*, the minimum utility threshold is set to be 5% while 10% is used as minimum utility threshold for datasets from *G_accidents*.

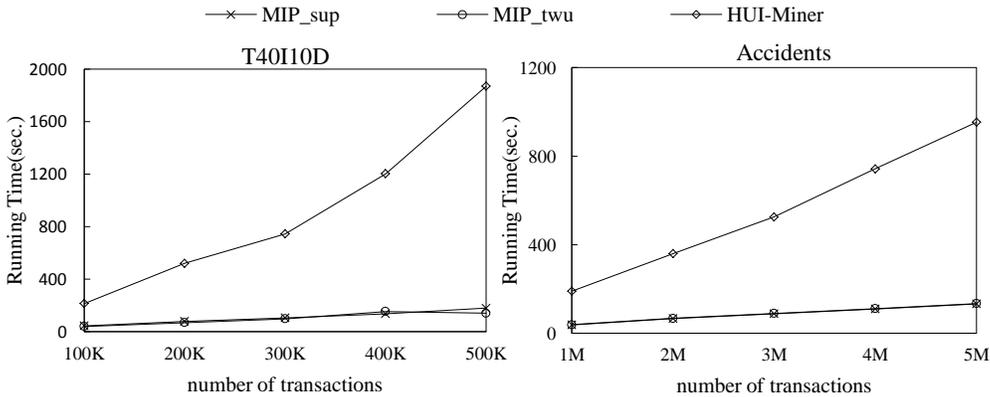

Figure 8: Scalability of Runtime with Number of Transactions

Figure 8 shows the speed scalability of MIP_sup, MIP_twu, and HUI-Miner on dataset group *G_T40I10* and *G_accidents*. In the figure, all algorithms increase linearly when the size of dataset increases. However, compared with HUI-Miner, MIP_sup and MIP_twu have better scalability. This can be explained as follows. According to the construction, all datasets in each group are built from the same probabilistic generative model. Therefore, the difference of data distributions among datasets in a group is small. As we know, if the distribution of data remains stable, the number of nodes in the PU-tree changes little when the size of dataset increases. Thus, the length of PUN-lists is sublinear with the number of transactions. However, the length of utility-lists is linear with the number of transactions. In addition, we observe that the curves for MIP_sup and MIP_twu overlap each other on both dataset groups, which means that they have almost the same scalability.

Figure 9 gives the memory scalability of the three algorithms on the two groups of dataset. The figure shows that memory consumption of all algorithms increases linearly when the size of datasets changes. MIP_sup and MIP_twu are better on dataset group *G_accidents* while HUI-Miner is better on dataset group *G_T40I10*. This can be explains as follows. Since each dataset in *G_T40I10* is constructed from the same data generator with the same parameter setting as *T10I4D100K*, their data distributions are similar to that of *T10I4D100K*. That is, they are sparse datasets. As discussed in Subsection 5.2, the performance of memory consumption of MIP is worse than that of HUI-Miner on sparse datasets. On the contrary, datasets in *G_accidents* are dense since they are sampled from *accidents*. According to the discussion in Subsection 5.2, MIP performs better on dense dataset than HUI-Miner in term of memory consumption. Therefore, MIP_sup and MIP_twu show better scalability.

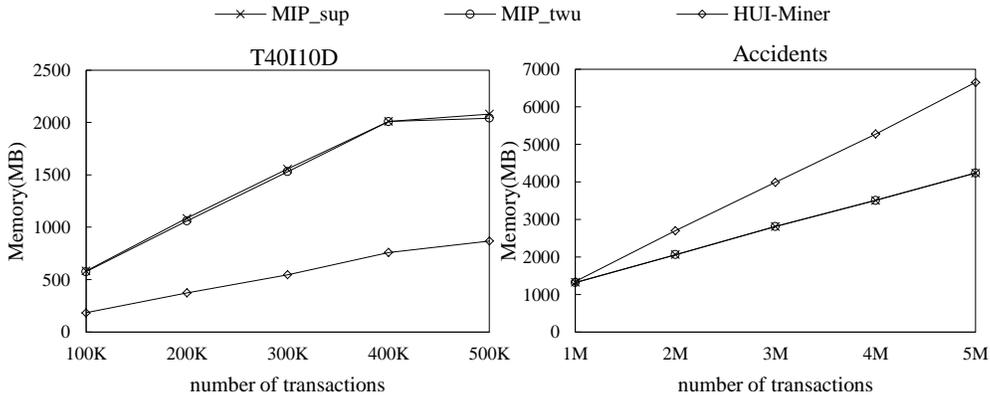

Figure 9: Scalability of Memory Consumption with Number of Transactions

In Figure 9, the curve for MIP_sup and MIP_twu overlap each other again on both dataset groups. This observation together with similar observations in Figure 6, 7, and 8 strongly indicates that item-support descending order and TWU descending order play almost the same role in mining high utility itemsets.

From the experimental results of runtime, memory consumption, and scalability, we can draw the conclusion that PUN-list is a very effective structure for mining high utility itemsets since MIP outperforms the state-of-the-art algorithms almost in all cases on both real and synthetic datasets. Generally, when the datasets are dense, MIP performs better. For a sparse dataset, its PU-tree is wide, sparse, and big, which makes the PUN-lists are not short enough. Therefore, the memory for MIP is high for sparse datasets. However, MIP is still the fastest algorithm and has very good scalability.

## 5 CONCLUSION

In this paper, we have proposed a novel data structure, PUN-list, which maintains the utility and utility upper bound information about an itemset by summarizing this information at different nodes of a PU-tree. Based on PUN-list, we developed an efficient algorithm, MIP, for high utility itemsets mining. Besides, MIP employs an efficient mining strategy that directly discovers high utility itemsets from the search space using pruning information stored in PUN-lists. We have studied the performance of MIP in comparison with the state-of-the-art algorithms on various real and synthetic datasets. The experimental results show that MIP outperforms these algorithms substantially, which strongly indicates that PUN-list is a very effective data structure for mining high utility itemsets.

Recently, there have been some emerging studies on mining top-k high utility itemsets [Wu et al. 2012], Concise and Lossless Representation of High Utility Itemsets [Wu et al. 2011; Wu et al. 2014] and High Utility Sequential Patterns [Wu et al. 2013; Yin et al. 2012; Yin et al. 2013]. The adoption or extension of PUN-list to mine these special high utility patterns is an interesting topic for future research. In addition, as the available data is growing exponentially, the parallel/distributed imple-mentation of our method is also an interesting work.